\documentclass[reprint, amsmath,amssymb, aps]{revtex4-1}
\usepackage[utf8]{inputenc}
\usepackage{amsmath}
\usepackage{amssymb}
\usepackage{hyperref}
\usepackage{graphicx}
\usepackage[dvipsnames]{xcolor}
\usepackage{comment}

\def\equationautorefname~#1\null{Eq.\,(#1)\null}
\def\figureautorefname~#1\null{Fig.\,#1\null}
\def\sectionautorefname~#1\null{Sec.\,#1\null}

\definecolor{crimson}{HTML}{DC143C}
\definecolor{cGreen}{HTML}{0DBB53}
\definecolor{bColor}{HTML}{B5077B}
\definecolor{GreyFirst}{HTML}{A502EB}
\definecolor{bracketColor}{HTML}{A10B22}
\definecolor{codeBlue}{HTML}{1809E8}

\date{\today}

\begin{document}
\title{Rare Events of Host Switching for Diseases using a SIR Model with Mutations}

\author{Yannick Feld}
\email{yannick.feld@uol.de}
\homepage{https://www.yfeld.de}
\author{Alexander K. Hartmann}
\affiliation{Institut f\"ur Physik, Carl von Ossietzky Universit\"at Oldenburg, 26111 Oldenburg, Germany}

\begin{abstract}
    We numerically study disease dynamics that lead to the disease 
switching from one host
species to another, resulting in diseases gaining the ability to 
infect, e.g.,  humans.
    Unlike previous studies that focused on branching processes starting 
with the first infected humans, 
    we begin by considering a disease pathogen 
that initially cannot infect humans. 
We model  the entire process, starting from an infection in the
animal population,    including mutations that eventually 
enable the disease to cause an epidemic outbreak
in the human population. We use an SIR model on a network consisting of
132 dog and 1320 human nodes, with a single parameter representing the gene
of the pathogen.

    We use numerical large-deviation techniques, 
specifically the $1/t$ Wang-Landau algorithm, 
 to calculate the potentially very small 
probability of the host switching event. With this approach we are able to resolve probabilities as small as $10^{-120}$. 
Additionally the $1/t$ Wang-Landau algorithm allows us to 
    obtain the complete probability density function $P(C)$ of the 
cumulative fraction $C$ of infected humans,    which is an indicator for 
the severity of the disease in the human population.

    We also calculate correlations of $C$ 
    with selected quantities $q$ that characterize the outbreak.
    Due to the application of the rare-event algorithm, this is possible
  for the entire range of $C$ values.

\end{abstract}

\maketitle

\section{Introduction}
Understanding the dynamics of disease transmission is a very important aspect for a variety of 
disciplines like immunology, biology, statistics, applied mathematics and statistical physics 
\cite{hethcote2000, salathe2010, apolloni2014, pastor-satorras2015, wang2017, walters2018, tang2020, handel2020, chen2021}.
Besides understanding the dynamics of existing diseases such research often aims to find effective strategies 
to fight disease outbreaks,
e.g., considerable efforts have gone into areas like non-pharmaceutical interventions \cite{brito2021}
or vaccinations \cite{wang2016}.

Clearly infectious diseases are not only a problem for humans but affect basically all high-level species on 
this planet and many pathogens are known to infect multiple hosts \cite{cleaveland2001}. 
In fact, most of the pathogens that infect humans are known to also infect at least one animal species \cite{taylor2001}. 
Despite that we only know a small fraction of viruses that affect domestic let alone wild animals \cite{Slingenbergh2004, Parrish2008, Luis2013, Guegan2020}.

On an evolutionary time scale we know of many pathogens that adapted to new hosts  \cite{holmes2022}, i.e.,
host-switching events. 
Such events pose a serious thread to human populations \cite{Parrish2008}
and there is a need to assess the risk that known diseases pose for spillovers.
This is achieved by considering several risk factors, like the number of contacts between 
humans and the current host species of the disease, which can then be combined into an overall risk factor \cite{grange2021}.

However, to get a more fundamental understanding of the underlying processes, there is a need to model these processes with more detail.
On the one hand this is quite useful to understand the importance of different risk factors better,
on the other hand it might provide better tools for quantifying 
said risks in the first place.

Models that calculate the probability of a disease switching to 
new host species, including mutation events such that the pathogen is able 
to cause an epidemic outbreak, have been studied before. 
Antia et al.~\cite{antia2003} used a multi-type branching processes and 
typical-event sampling methods for their study.
A similar approach was used to calculate the spillover probability and other 
quantities for coupled metapopulations \cite{singh2014ZOONOTIC}.
Another study started with a single infected human and used a branching-process to estimate the probability that 
avian influenza mutates and becomes a pandemic in the human population \cite{tripathi2021}.
Furthermore, Schreiber et al.~\cite{Schreiber2021} investigated the evolutionary emergence of infectious 
diseases with a combination of within-host dynamics, 
which explicitly model the viral load in infected individuals, and a multi-type branching process on the population level. 
Spillover from a reservoir community was also investigated with steady-state analysis \cite{ng2023}.

Still, most of these studies begin with the host-switching event, i.e., they start with the first human infection. This means that
pathogens that correspond to subcritical Zoonoses, i.e., have a very small probability of infecting humans and would have to mutate 
to cause larger outbreaks in the human society, 
are currently not  studied much \cite{mummah2020}. Thus,
the animal-human interface is mostly overlooked in modelling and 
the host switch itself
is rarely incorporated \cite{LloydSmith2009,dorjee2013}.
Also, as Antia et al.~\cite{antia2003} mentioned, 
it would make sense to include 
the genetic diversity of the pathogen in the animal population.

For studying the probability of a new host-switching event, processes 
that lead to a very high switch
probability are of limited relevance in practical terms, because such switches 
are likely to have already occurred during evolution. Thus,  we consider
diseases here, where the switch has not taken place yet.

We should, first, note that more than $10^9$ 
humans are constantly exposed to many foreign pathogens that 
could potentially infect them.
Considering the large number of more than a $10^{12}$ 
microorganisms on the human skin 
alone \cite{ursell2012,byrd2018},
not to mention the abundance of microbes on the entire planet 
\cite{Kallmeyer2012, flemming2019}, it is clear that 
contacts of humans with potential pathogens are frequent. But since
Zoonotic spillovers are relatively rare events \cite{Plowright2017},
 the probability 
of any \emph{single} pathogen gaining the ability to infect humans 
is actually very small. Alone the number humans times the number
of microorganisms per human results in $10^{-21}$ ongoing contacts,
so the overall spillover probability per contact must be much smaller,
which makes it hard to treat the process by 
simulations with standard approaches.

On the other hand, given the large total number of human-animals contacts,
 it becomes likely that occasionally 
one of those diseases manages to perform a host switch.
The emergence of COVID-19 \cite{berber2021, holmes2022} 
serves as a recent example
and there are many other examples of cross-species transmissions that also caused serious harm \cite{Daszak2000,Daszak2004}. Thus, it is very relevant
to model such host-switching processes in particular in the 
regime of very small switching probabilities, to be able to, at least
in principle, estimate the risks better.

Thus, in this study, we present a numerical rare-event 
study of such switches using a variant \cite{rudiger2020} of the well-known SIR model \cite{Kermack1927, kiss2016, brauer2019},
which incorporates mutations.
Simulating very rare events poses challenges, since typical-event sampling methods 
are not feasible due to the high amount of computational power they would require.
The need for studying rare events is, however, not exclusive to disease 
dynamics but also 
important for a variety of other areas and has been performed
 by numerical \cite{dembo2010,giardina2011}
and analytical or mathematical approaches \cite{denHollander2000,touchette2009,touchette2011}.

Recently the authors of the current work 
have applied \cite{feld22} large-deviation algorithms to the standard 
SIR model  without mutations and for a single species.
In the current study we build upon those previous work, extend it to two species and 
incorporate mutations,
which allows us to study cross-species transmissions with high numerical 
precision to calculate switching probabilities even as low as $10^{-120}$.
We also explore correlation patterns with other measurable quantities, 
which further 
enhances our understanding of cross-species transmission and transmission 
of mutating  diseases in general.

The remainder of this paper is structured as follows: First we introduce the SIR model, followed 
by the presentation of the utilized network model. We explain the used large-deviation techniques
and provide a small simple sample study where we explore the parameter space before presenting 
the results of our large-deviation investigation. 
Finally, we give a summery and an outlook.

We believe that this study will contribute to the growing body of knowledge in disease 
transmission dynamics and provide, on an abstract level,
first valuable insights into 
the risk of cross-species transmission events.

\section{SIR model \label{sir_model_sec}}

We extend an SIR model that was modified to incorporate mutations \cite{rudiger2020} as it is explained below.

Each node of a given network is in either of three states \emph{Susceptible} ($S$),
\emph{Infected} ($I$) or \emph{Recovered} ($R$). The model is defined by a 
global \emph{recovery probability} $\mu$, here we use a value $\mu=0.14$ for
simplicity, which is somehow arbitrary 
since it basically just fixes the time scale. For details on the dynamics 
of the SIR model, see  below.

Additionally, for each infected node $i$ a, for simplicity single-valued,
 gene variable $\gamma_i \in \mathbb{R}$ is stored, which is utilized 
to determine the  transmission probability $\lambda_i$ of the corresponding 
pathogen hosted by node $i$.
In the original paper \cite{rudiger2020} the transmission probability $\lambda$ was defined to be 
some function $\lambda(\gamma)$. This is a very simple representation
of a fitness landscape. At least $\lambda(\gamma)$ should exhibit a maximum
representing the variant of the pathogen which transmits best.

We take a similar approach in our study;
however, we aim to investigate a disease that switches from one host 
species to the next.
Consequently, each infected node is associated with two lambda values, i.e.,
the transmission probability to animals $\lambda^{a}_i$ and the 
transmission probability to humans $\lambda^{h}_i$. Thus, for simplicity,
the transmission probability depends only on the target species and
on the gene value $\gamma$, not on the current species.
Consequently  every individual, regardless whether it is 
an animal or 
a human, exhibits both transmission probabilities because the corresponding
pathogen might be transmissible for both species.

As in the reference work \cite{rudiger2020}
we assume that the fitness landscape exhibits, beyond the most
simple case, more than a single
maximum for the infection probability to, e.g., account for different routes
of transmission. For this we used the following function
modelling  the transmission probability to animals:
\begin{equation}
    \lambda^a(\gamma) = \lambda^{\text{max}} \frac{ \left(2-\gamma^2\right) \left(\cos\left(5 \gamma\right)+2\right)}{6}\,.
\end{equation}
Here $\lambda^{\text{max}}$ represents a parameter that defines the maximal transmission probability 
that the disease can potentially reach.
In the context of this work we always use $\lambda^{\text{max}}=0.15$. Together with the value of $\mu=0.14$ 
this means that a global disease is possible.

Mutations that increase the ability of a disease to infect a new host species
are likely to decrease the ability of said disease to infect the old host species \cite{Parrish2008}.
However, while we did not want the functions $\lambda^a(\gamma)$ and $\lambda^h(\gamma)$ to be the same,
we also did not want them to differ too much.
Thus, we decided to use the same function shape 
for both but  slightly shifted. There needs to be an overlapping region,
where the transmission probability is nonzero for animals and humans,
to allow for an evolution of the gen variable $\gamma$. Here, we chose 
that the point where $\lambda^h(\gamma)$ starts to differ from 0 aligns
 with the point where the transmission probability for the animals exhibits a 
local but not a global maximum, specifically at $\gamma_m=1.00728…$, i.e.,
\begin{align}
    \gamma^\prime &= \gamma - 2.4214957… \approx \gamma - \sqrt{2} - 1~,\\
    \lambda^h(\gamma) &= \lambda^a(\gamma^\prime)~.
    \label{EQlambdaH}
\end{align}

\begin{figure}[htb]
    \centering
    \includegraphics[width=\linewidth]{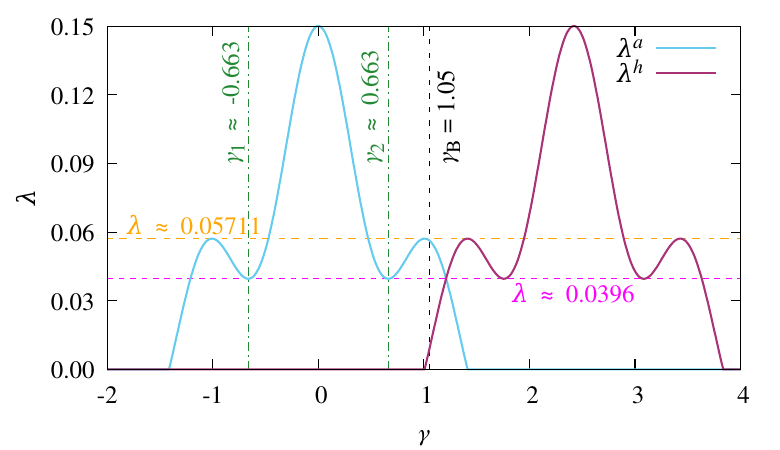}
    \caption{Functional relation of the gene variable $\gamma$ and the transmission probability to humans $\lambda^h$
    in red and the transmission probability to animals $\lambda^a$ in blue. 
    We have also highlighted the positions of the two local minima $\gamma_1$ and $\gamma_2$ with dashed vertical lines. The highlighted value 
    of $\gamma_B=1.05$ will become important later in this paper and is thus also included.
    The height of the local minima and maxima is highlighted with the dashed horizontal lines.}
    \label{fig:func}
\end{figure}

Note that it can be assumed that the qualitative behavior of the
model will not depend on the actual shapes and relative
weights of the functions.
In \autoref{fig:func} we show the functions we used to calculate $\lambda$ 
from $\gamma$.
For clarity we highlighted the most important values in the plot.

Having defined the transition probabilities, we can now proceed to 
the actual dynamics.
To initiate a SIR simulation, all nodes are assigned the $S$ state, except for one randomly selected 
node from the animal network, which we also call \emph{patient zero}. This node is set to the infected state $I$ and assigned the initial gene value 
$\gamma=\gamma_{\text{init}}$.

To perform a time step we iterate over all susceptible nodes that are adjacent to at least one infected node in a parallel fashion.
Let us consider node $j$ with infected neighbors $i$, each having corresponding gene values denoted by $\gamma_i$.
Accordingly, the node $j$ will be flagged to become infected in the next
time step with a probability of 
\begin{equation}
    \lambda_j = 1 - \prod_{i} (1-\lambda(\gamma_i))\,.
    \label{lambda_i_eq}
\end{equation} 
If the node is flagged we need to determine which specific node $i$ actually caused the infection of  
$j$. This is done by drawing one node $i_0$ 
from all infected neighbors such that
 each node $i$ exhibits a probability of being chosen 
that is proportional to its corresponding $\lambda(\gamma_i)$ value.
Next, we assign the  value of the gene variable 
\begin{equation}
    \label{gamma_eq}
    \gamma_j = \gamma_{i_0} + \varphi
\end{equation}
where $\varphi$ is a random number drawn from a Gaussian distribution with a mean of 0 and a variance of $\sigma$.
Note that $\sigma$ can be understood as a sort of mutation rate, i.e., a low value of $\sigma$ will typically result in 
only minor mutations whereas a large value of $\sigma$ lets large mutations 
appear more frequently.

Now that all infections for the next time step have been decided, we need to decide on the recoveries.
For this we iterate over all infected nodes and transition them to the recovered state $R$ with probability $\mu$.
To conclude the time step, we update the state of all flagged nodes to $I$.

It is worth noting that we implemented a slightly different algorithm 
than the one used in Ref.~\cite{rudiger2020}, 
but note that both implementations are equivalent.
This was necessary for algorithmic reasons to incorporate
the large-deviation simulations which we describe
 later on. For this purpose we, rather than drawing numbers on demand, 
pre-generate and store them in vectors,
such that they can be manipulated in a controlled fashion.
More details are provided in \autoref{sec_ld} and 
\autoref{MCMC_MOVE_section}. 
For this it is beneficial to reduce the amount of required random numbers, 
which is the reason for the different implementation.

Since, for every infected or recovered node $j$, we have the information about which node $i_0$ was responsible for the infection, 
we can construct an \emph{outbreak tree}.  
In this tree the initial infected node, i.e., patient zero, serves as root. 
Directional edges are 
created from each node to the nodes it infected.
Nodes that were never infected are disregarded. 
The resulting outbreak-tree represents a subgraph of the original graph.

Lastly we need a quantity that characterizes the severity of the outbreak in the human 
population. For this we use the cumulative fraction of humans that contracted the disease 
during the outbreak, which we will denote by $C$.

\section{Network ensemble \label{sec:network_ensemble}}

Since we aim to model a disease that switches from one host species to another we now need to model 
two different host species with some links in between.

Considering that this study is fundamental research, rather than a study 
tailored for a specific disease or location, 
the details of the contact networks should not matter much. 
We still wanted to investigate 
a relatively realistic case and chose to use a network model \cite{Laager2018}
that was fitted to a population of dogs in  N'Djam\'ena, Chad,
in order to evaluate measures against rabies. 
To model the network the authors of  have 
measured the contact network of wild and domesticated dogs and have 
fitted a spacial \cite{barthelemy2022} network model to the data.
The construction of this resulting spacial network works as follows:

To create a network of $N_a$ animal nodes we first have to decide their x- and y-coordinates,
which should be located in the unit square. 
The coordinates are chosen using Latin Hypercube sampling \cite{McKay1979, tang1993}, i.e.,
the x-coordinate is sampled exactly once from every interval 
$\left\{\left[0, \frac{1}{N_a}\right), \left[\frac{1}{N_a}, \frac{2}{N_a}\right), …, \left[\frac{N_a-1}{N_a}, 1 \right)\right\}$
in random order. The y-coordinate is sampled in the same way.

Next we iterate over every pair $i<j$ of nodes and connect them with 
the probability 
\begin{equation}
    \label{inc_prob}
    p_{i,j} = 1-\left(1-\left(e^{-\kappa \sqrt{(x_i-x_j)^2+(y_i-y_j)^2}}\right)\right)^2\,,
\end{equation}
which depends on the euclidian distance of the nodes and some scaling variable $\kappa$, the value that was actualy used is listed below.

Additional to those spatially motivated edges, a fraction of nodes are now selected to become 
something akin to hubs by connecting them to additional nodes without any regard for spacial distance.
For this we randomly draw $k=N (1-\tau)$ nodes (rounded to the closest integer), 
where $\tau$ is a second model parameter.

For each node $i$ in the hub-set we first decide the number $m$ of hub-links we want to create by
drawing $m$ from a Poisson distribution with mean $\Lambda$, 
where $\Lambda$ is the last model parameter.
Then we iteratively add $m$ new edges to the node $i$ where the probability of connecting it to 
node $j$ is 
\begin{equation}
    \tilde{p}_{i,j}  = \frac{k_j}{\sum_{l=0}^{N-1} k_l}\,,
\end{equation}
where $k_j$ is the current degree of node $j$.

Throughout this work we used $\Lambda=7$, $\tau=0.7$ and $\kappa=10 \sqrt{N_a/66}$
for the parameters, which were taken from \cite{Laager2018}, although we 
scaled $\kappa$ such that we can use this parameter set for a different number $N_a$ of animals.

This finalizes the animal part of the network. Next we construct a network intended to 
represent the human population. For this we chose to use a \emph{small-world network} \cite{watts1998}.
While this does not perfectly describe human contact networks it is a good-enough approximation for the 
purpose of this study.

The network of humans is initalized with $N_h$ nodes 
$i=1,\ldots, N_h$.
First, every node $i$ is connected to all neighbors $j$ for which 
$|i-j| \leq 8$ mod $N_h$, i.e., with periodic boundary conditions. This  
creates a ring in which every node now has an initial degree of 16.
Next we iterate over all edges $\{i,j\}$ once and rewire each with probability $p$ 
to a random node $j^\prime \neq i$, i.e., we swap $\{i, j\} \to \{i, j^\prime\}$.
We use a rewiring probability of $p=0.1$ throughout this paper.  

This completes the construction of the separate contact networks for each host-species.
Now we still need to create some edges between the networks, i.e., connect a few animals with humans. 

The fraction of ownerless dogs was estimated to be between 8-15\% \cite{lechenne2016, Laager2018}. Here we use 15\% ownerless dogs. 
For all other dogs we each drew an 
owner from the set of humans that not yet owned a dog and then we created an edge between 
them. Thus, all dogs with an owner now have exactly one edge connecting them 
to a human node.

\section{Large deviations \label{sec_ld}}

To be able to calculate very small probabilities we need to employ 
special large-deviation 
algorithms. Under the name of \emph{transition-path sampling} \cite{crooks2001, dellago1998} 
these methods gained their initial popularity in statistical physics.
Since then these large-deviation algorithms have been applied to a variety of models, including
but not limited to power grids \cite{Dewenter_2015, feld19}, 
the Kardar-Parisi-Zhang equation \cite{Fogedby2009, hartmann2019, hartmann2021},
Ising models \cite{Koerner2006, Fytas2012, Pommerenck2020, Li2022}
as well as to measure 
various graph  \cite{hartmann2004, hartmann2011, hartmann2017, schawe2019},
RNA \cite{Werner2021} and protein
properties \cite{Ojeda2009, Swetnam2011, Singh2014}.

For applying these methods to the SIR model, the large-deviation simulation 
needs to be able to control the underlying SIR simulations \cite{feld22}, i.e., 
the SIR dynamics need to be manipulated in a controlled fashion.
This allows one to focus on different, originally rare, parts of the
dynamics. Since the control is known, one can easily obtain the true
extremely small probabilities of the observed
 events during the subsequent analysis of the results.

This is done as follows:
In a standard SIR simulation, random numbers uniformly distributed in the interval $\left[0,1\right]$ are typically generated on demand.
By comparing those numbers against the respective transmission probability $\lambda_i$ (see \autoref{lambda_i_eq}) or the recovery probability $\mu$ 
one can decide whether a susceptible node becomes infected or an infected node becomes recovered.
In case of the SIR model with mutations that is applied in this work, 
we need to make additional random choices once a new node $i$ is infected:
One uniformly distributed random number is required to decide which of 
$i$'s infected neighbors caused the infection, only
when node $i$ has one infected neighbor this step can be skipped.
This information is required to decide the respective $\gamma_i$ 
value according to \autoref{gamma_eq},
for which we also need a Gaussian distributed random number $\varphi$.

Instead of drawing the random numbers on the fly one could create them beforehand and store them in the vectors $\xi^{a}_\mu$, $\xi^{h}_\mu$, 
$\xi^{a}_\lambda$, $\xi^{h}_\lambda$,
$\xi_\theta$ and $\xi_\sigma$. $\xi^{a}_\mu$ ($\xi^{h}_\mu$) 
and $\xi^{a}_\lambda$ ($\xi^{h}_\lambda$) contain a distinct random number 
for each animal (human) and time step $\tau$. The number 
of random numbers in $\xi_\theta$ and $\xi_\sigma$ is independent of the 
number of time steps and only dependent on the number of nodes,
because each node can be infected at most once 
(see \autoref{MCMC_MOVE_section} for more details).
Now $\xi_\mu$ ($\xi_\lambda$) can be used to decide if a node $i$ becomes recovered (infected) at a given time step.
Upon infection $\xi_\theta$ is used to decide which neighbor of node $i$ infected it such that finally $\xi_\sigma$ can be used to 
decide the respective value of $\gamma_i$.
Lastly, we store the index of the first infected node in the variable $\xi_0$, which is a number uniformly drawn from all indices corresponding
to animals.

As long as the length of the random number vectors are long enough such that the disease outbreak terminates before 
the simulation runs out of random numbers, this procedure cannot change the outcome of the simulations.
However, the entire outbreak simulation is now a deterministic outcome of the randomness contained within 
$\Xi = \left(\xi^{a}_\lambda, \xi^{h}_\lambda, \xi^{a}_\mu, \xi^{h}_\mu, \xi_\theta, \xi_\sigma, \xi_0\right)$.
Note that random numbers are occasionally ignored, e.g., random numbers corresponding to nodes that, at that time step, have already 
recovered will have no effect.

One could do \emph{simple sampling} if one drew 
independent vectors $\Xi$ many times,
for each of which any desired quantity, here $C$,  
would be evaluated to create a histogram. 
This would enable one to estimate the high-probability part of the
distribution $P(C)$. To go beyond this and estimate
the distribution over a large range of the support, we use
this setup to sample rare events by controlling the values within $\Xi$ via a Markov-Chain-Monte-Carlo (MCMC) approach.
For this purpose, 
we employ the $1/t$ \emph{Wang-Landau} algorithm \cite{Belardinelli-2007}, which is an improved version of the original 
\emph{Wang-Landau} \cite{WangLandau-2001} algorithm that prevents error saturation \cite{yan2003, Belardinelli-2007, Belardinelli-2008, Belardinelli2016}.

The WL algorithm requires an initial estimate $h(C)$ for the probability density distribution $P(C)$.
This estimate does not need to be normalized and
it is usually sufficient to use, e.g., $h(C)=1~\forall C$,
though if one has prior information about the pdf one can, of course, supply a better estimate.

Now a Markov-Chain with the steps $t=0,1,…$ is created.
For each step $t$ a new \emph{trial configuration} $\tilde{\Xi}$ is 
constructed based on the previous
configuration $\Xi^{(t-1)}$ via the 
Markov moves explained in \autoref{MCMC_MOVE_section}.
Each of those configurations deterministically determine an entire outbreak simulation
and thus correspond to the resulting cumulative fractions $\tilde{C}$ and $C^{(t-1)}$,
that can be calculated by performing the respective simulations.

To decide whether to accept, i.e., $\Xi^{(t)}=\tilde{\Xi}$, or reject, i.e., $\Xi^{(t)} = \Xi^{(t-1)}$,
the trial configuration, the Metropolis-Hastings\cite{Hastings1970} probability 
\begin{equation}
    p=\min\left[1,\frac{h(C^{(t-1)})}{h(\tilde{C})}\right]
\label{eq:metropolis}
\end{equation}
is used. This means that the acceptance probability for the WL algorithm
is inverse proportional to the
current estimate $h(C)$ of the probability density function.

To refine the probability density estimate $h(C)$, WL uses a multiplicative 
factor $f>0$. It is utilized in each step $t$ to 
changing the estimate as $h(C^{(t)}) \to f h(C^{t})$, 
while leaving the estimate for the values of $C$ untouched. Thus,
if the simulations remain at some value of $C$ for a while, it will
 subsequently become less and less likely
to further remain there due to \autoref{eq:metropolis}.

In the beginning the factor is usually relatively large, e.g., $f=e=2.71…$ and then progressively reduced 
via some schedule. This gradual reduction allows the estimate $h(C)$ to be updated on a finer and finer scale,
such that, apart from the normalization, it ultimately converges to the sought-after pdf $P(C)$.
Thus, we can obtain 
\begin{equation}
    P(C)=\frac{h(C)}{\sum_C h(C)}~.
\end{equation}
For details about the schedule for changing $f$  we refer to the 
literature, just keep in mind that the update schedule is actually the 
main difference between the original WL algorithm \cite{WangLandau-2001}
and the $1/t$ WL algorithm \cite{Belardinelli-2007}.

The convergence properties of the Markov chain depend on the chosen
parameters of the model.
The described algorithm works well when we start with an initial $\gamma$ value close to 1,
i.e., where it is quite likely that the disease switches from the animals to the humans.

If, however, we have a low value of gamma, e.g., $\gamma=-0.663$, and a 
low mutation rate, i.e., a low value of $\sigma$, 
then we experience some issues:
Let us consider a MCMC chain that currently
exhibits the configuration $\Xi$ where the disease does not switch to the humans,
i.e. one has $C=0$.
It now becomes very hard to escape, as there is likely not one single 
Markov step that 
can change the configuration to a state where the humans become infected,
i.e. $C>0$.
Instead, a sequence of relatively specific Markov moves would be required.
From the perspective of the simulation, however, the intermediate 
configurations all 
correspond to $C=0$ and thus they all correspond to the same bin in the 
histogram. 

This is an issue, because all moves that do not change the bin will 
be accepted since the 
corresponding Metropolis-Hastings probability becomes 1.
As a result the Markov moves will randomly move in the configuration space that 
corresponds to $C=0$, without any ``drift'' towards $C>0$.
Only after a long time it might manage to randomly switch to $C>0$. 
This means the estimate of $h(C=0)$ might have grown to a high value.
Thus, a move that leads back to $C=0$ is very unlikely to be accepted,
at least for some time. This is bad, because for a good
convergence, the WL algorithm should visit all possible bins frequently.
Furthermore, we observed that within the simulations the first infected 
human  does not change anymore, which means 
that we are also restricted to a configuration subspace and 
thus have issues with \emph{ergodicity} \cite{newman1999, landau2014}.

Since we could pinpoint our issues to the $C=0$ bin of the histogram 
we were able to solve them by a quasi two-dimensional
histogram indexed by
$C$ and $\gamma_{\max}$, where the latter is the current maximum 
of $\gamma$ encountered in the animal population
during an outbreak defined by the current randomness $\Xi$.

With respect to the shape of $\lambda(\gamma)$ we distinguish values
of $\gamma_{\max}$ as follows: Firstly, values smaller than 
$\gamma_{\text{init}}$, which is the value of patient zero, are as considered as similar, i.e., lumped together. 
Secondly, 
values in between $\gamma_{\text{init}}$ and 1.05 are most important. 
Thus, this interval is subdivided
into $\hat{N}$ sub intervals, where $\hat N$ can be chosen somehow
arbitrarily,
here we used values in the range $\hat N\in \{100,…,1100\}$, depending on the chosen value of $\gamma_{\text{init}}$.
 Thirdly, all
values $\gamma_{\max}>1.05$ are also lumped together.

Instead of storing a full two dimensional histogram $h(C,\gamma_{\max})$,
we map it to a one-dimensional one. For this purpose
let us first define the binning for the values of 
$\gamma_{\max} \in [\gamma_{\text{init}},1.05]$:
\begin{align}
    b(\gamma_{\max}) &= \left\lfloor \hat{N}\left(\frac{\gamma_{\max} - 1.05}{1.05-\gamma_\text{init}}\right) \right\rfloor\,, 
\end{align}
which is smaller than zero.
We define a new quantity $\Psi$, which denotes the index in the
one-dimensional histogram for encountered values ($C,\gamma_{\max}$),  as 
\begin{equation}
    \Psi(C, \gamma_{\max}) = \begin{cases}
        C &\text{if } C > 0\\
        0 &\text{else if}~b(\gamma_{\max}) \geq 0\\
        -\hat{N} & \text{else if}~b(\gamma_{\max}) \leq -\hat{N}\\
        b(\gamma_{\max}) & \text{otherwise}  
    \end{cases}\,,
\end{equation}
Thus, the index ranges from $-\hat N$ to the number $N_h$ of humans.

Sometimes we are only interested in the probability that the disease switches to the human population.
In that case we do not care about the actual size $C$ of the outbreak and 
thus we can use a single bin to 
account for all values $C>0$, which reduces the required computation time.

By now calculating the non-normalized probability estimate $h(\Psi)$ via WL analog to what is described above we are able to actually sample the pdf,
as now we are able to reach $C>0$ from bins with high $\gamma_{\max}$ values. 
We can normalize $h(\Psi)$ such that the sum of all bins equals 1, i.e., $\sum_\Psi h(\Psi)=1$. 
Note that we can recover $h(C=0)$ via
\begin{equation}
    h(C=0)=\sum_{\Psi=-\hat{N}}^0 h(\Psi) = 1- \sum_{\Psi>0} h(\Psi)\,,
\end{equation}
where the latter equality holds due to the normalization.

This approach enables the sampling of extremely rare events that cannot be accessed through 
typical-event sampling (also known as simple sampling) methods. 
As a result, it allows for the sampling of distinctive features of the pdf across its entire support.

Strictly speaking WL does not fulfill \emph{detailed balance} \cite{newman1999}, however,
since $h(\Psi)$ is continuously updated.
To address this  we additionally apply \emph{entropic sampling} \cite{Lee-1993},
which is very similar to WL, it just does not update the estimate $h(C)$ 
of the pdf during the simulation but only 
updates it afterwards.
This step was here not essential for estimating $P(C)$, as the accuracy 
achieved 
by WL turned out to be already exceptionally high, 
making the subsequent entropic sampling calculation only marginally beneficial.

Nonetheless, the additional entropic sampling simulation enabled us to achieve a rather uniform sampling of disease trajectories 
across the entire range of possible $C$ values, which in turn allowed
 us to calculate correlation with other 
measurable quantities, even in the range of very improbable values of $C$. 

All in all, this rigorous numerical method provides high confidence 
in the results and  has proven to be very fruitful in the past.
\section{MCMC Moves \label{MCMC_MOVE_section}}

In this section we will show how the trial configuration $\tilde{\Xi}$ is created by making 
small changes to a given current configuration $\Xi^{(t)}$. 
Since the different vectors in $\Xi$
influence the disease dynamics in a different way, we need several
types of \emph{moves}. We first explain the special ones.

With a probability of $1\%$ a \emph{rotation move} is performed.
The rotation move is split into three sub-moves, the human-rotation, the animal-rotation and the combined rotation,
 of which one is randomly and uniformly selected.
For the human rotation $\xi^h_\mu$ and $\xi^h_\mu$ are rotated by $N_h$ to the left (50\%) or right (otherwise).
Similarly, for the animal rotation $\xi^a_\mu$ and $\xi^a_\mu$ are rotated by $N_a$ to the left (50\%) or right (otherwise).
The combined rotation works by rotating $\xi^h_\mu$ and $\xi^h_\mu$ by $N_h$ and $\xi^a_\mu$ and $\xi^a_\mu$ by $N_a$ to the left (50\%) or right (otherwise).
Those rotations roughly correspond to shifting the underlying time series by one time step to the left or right.
Note that, instead of copying a lot of RAM around, it is more efficient to just store the current rotation offset.

Before explaining the \emph{mutation change moves} we first need to clarify a technicality.
The vector $\xi_\sigma$ does not contain Gaussian distributed random numbers, but random numbers uniformly distributed on the
interval $(0,1]$ instead. Using the Box-Muller method \cite{box1958, hartmann_practical_guide2015} we can transform two uniformly distributed 
random numbers $u_1, u_2$ into two independent normal-distributed random numbers $\tilde{n}_1, \tilde{n}_2$.
We opted to always use only $\tilde{n}_1$, even though $\tilde{n}_1$ and $\tilde{n}_2$ are uncorrelated and one could 
technically use both.
We do this, because both random numbers $\tilde{n}_1,\tilde{n}_2$ 
would change upon changing one of the input random numbers $u_1,u_2$, but we 
want the simulation to be able to easily change single random numbers without automatically changing another.
This gives the simulation a finer control over the mutation changes.

The vector $\xi_\sigma$ contains $2 (N_h+3 N_a)$ random numbers uniformly drawn from $(0, 1]$,
which correspond to $N_h + 3 N_a$ Gaussian distributed random numbers.

The first $N_h$ random numbers are used to calculate the new gene value $\gamma$ if a human gets infected by another human,
the next $N_a$ 
 values are used if an animal infects a human, 
the next $N_a$ values are used if an animal gets infected 
by a human and the last $N_a$ values are used if an animal 
gets infected by another animal.
Note that these entries are not used randomly, but there exists a 
mapping, i.e., 
which entry we use depends on the index of the node in question.
Now that this is clarified we will come back to the change moves.

With a probability of $3.5\%$ we perform a \emph{simple mutation move}.
For this we repeat the following between 1 and 22 times (uniformly distributed). 
 uniformly draw an index of $\xi_\sigma$,
corresponding to a pair of two uniform random numbers
and exchange these numbers with newly drawn ones, 
which is equal to drawing a new sample from the 
Gaussian distribution. 

Note that changes of entries corresponding to nodes that, 
given $\Xi$, are not getting infected at all, 
will not have any effect on the simulation and thus will be accepted
by the Metropolis criterion. 
Also changes that correspond to leafs in the 
current outbreak-tree
will likely have a smaller effect and thus also have a high probability of 
getting accepted.
This results in a high over all acceptance rate of this type of move.
In Contrast, changes that effect the
children (in the outbreak-tree) of the initial infected node or more generally nodes on the path to the first infected human 
have a high likelihood of getting rejected, which results in a less efficient sampling.

To combat that, we introduce the \emph{tiny mutation change move}, which is performed with a probability of $3.5\%$.
This move is the reason why we use uniformly distributed random numbers followed by the Box-Muller method 
instead of directly using random numbers from the normal-distribution. 
Having uniformly distributed numbers lets us apply an idea that was first used in Ref.~\cite{Schawe2018}:
Instead of redrawing the pair uniformly distributed numbers 
$u^i_1, u^i_2$ corresponding to the $i$th index of $\tilde{\xi}_\sigma$,
we can just change them slightly, i.e., 
$\hat{u}^i_{1,2} = u^i_{1,2} + \chi_{1,2} \varepsilon$, where $\chi_{1,2}$  is 
uniformly distributed in $[-1,1]$ and $\varepsilon$ is uniformly drawn from the set
$\varepsilon \in \left\{10^{-i} | i \in \{0,1,2,3,4,5,6,7\}\right\}$.
If the resulting number is outside the allowed range, i.e., if $\hat{u}^i_{1,2} \notin (0,1]$,
then value is rejected, i.e., $ \hat{u}^i_{1,2} \to u^i_{1,2} $,
which is necessary to assure that the resulting values $\hat{u}^i_{1,2}$ are also distributed 
according to the correct uniform distribution. 

Now, if a \emph{tiny mutation change move} is selected, we do the following between 2 and 44 times (uniformly distributed).
First draw a random index $i$.
Then either ($66.6\%$) do the above process to only one of the uniform numbers, i.e., either to $u^i_1$ or to $u^i_2$, or ($33.3\%$) to both random numbers,
using the same $\varepsilon$ value for both random numbers but different
relative shifts $\chi_{1,2}$.
Overall the \emph{tiny mutation change move} was found to greatly improve convergence.

With a probability of 1\% a \emph{decision move} is performed, i.e., 
we perform the following 132 times:
Uniformly draw a random index $i$ of $\xi_\theta$. Exchange the $i$th 
entry of $\xi_\theta$ with a new random value, uniformly drawn from $[0,1]$.

With a probability of 1\% we perform a \emph{focused time move} which
changes some of the random number determining the initial phase of an outbreak.
For this, we first draw a random number $\omega$ uniformly from $\{0,1,…,29,30\}$. 
Then we redraw all random numbers within $\xi^h_\lambda, \xi^a_\lambda, \xi^h_\mu$ and $\xi^a_\mu$ 
that are associated with the $\omega$'th time step.

With a probability of 1\% a \emph{patient move} is performed by uniformly drawing a new animal index for the initial patient $\xi_0$.

Lastly, if none of the other moves was selected, i.e., with probability 89\%, we perform a \emph{randomize dynamics move} by doing the 
following 2100 times:
Select a random entry $\chi$ of $\xi^h_\lambda$, $\xi^h_\mu$, $\xi^a_\lambda$ or $\xi^a_\mu$ in such a way that every 
entry has the same probability of being chosen.
Then draw a uniformly distributed random number $u\in[0,1]$ and set $\chi \to u$.

Note that we document our move choices here for completeness reasons and to make it easy to reproduce the results.  
The correctness of the algorithm does not depend on the exact choice of moves or their exact 
relative frequency, as long as ergodicity is fulfilled.
It will, however, affect the efficiency of the algorithm and the speed of convergence.
As a rule of thumb one aims for an acceptance rate of about $\sim50\%$.
We have chosen the relative frequencies
determining which type of move is selected and the number of changes
perform to the corresponding entries by some experiments with this rule
of thumb in mind but by no means we have performed an exhaustive
simulation parameter test series.

\section{Simple sampling}

Next we wanted to sample the model parameter space. 
As explained earlier we always use a recovery probability of $\mu=0.14$, while the maximal transmission 
was limited to $\lambda_{\max}=0.15$.

We created a graph with $N_a=132$ dogs and $N_h=1320$ humans, which we will henceforth use for all simulations.
To scan the parameter space we considered 
200 values for the initial value of $\gamma_{\text{init}}$ evenly spread out in the interval $[-2,4]$ 
and 200 values for the mutation rate $\sigma$ that were evenly spread in the range $[0,10]$.

\begin{figure}[htb]
    \centering
    \includegraphics[width=\linewidth]{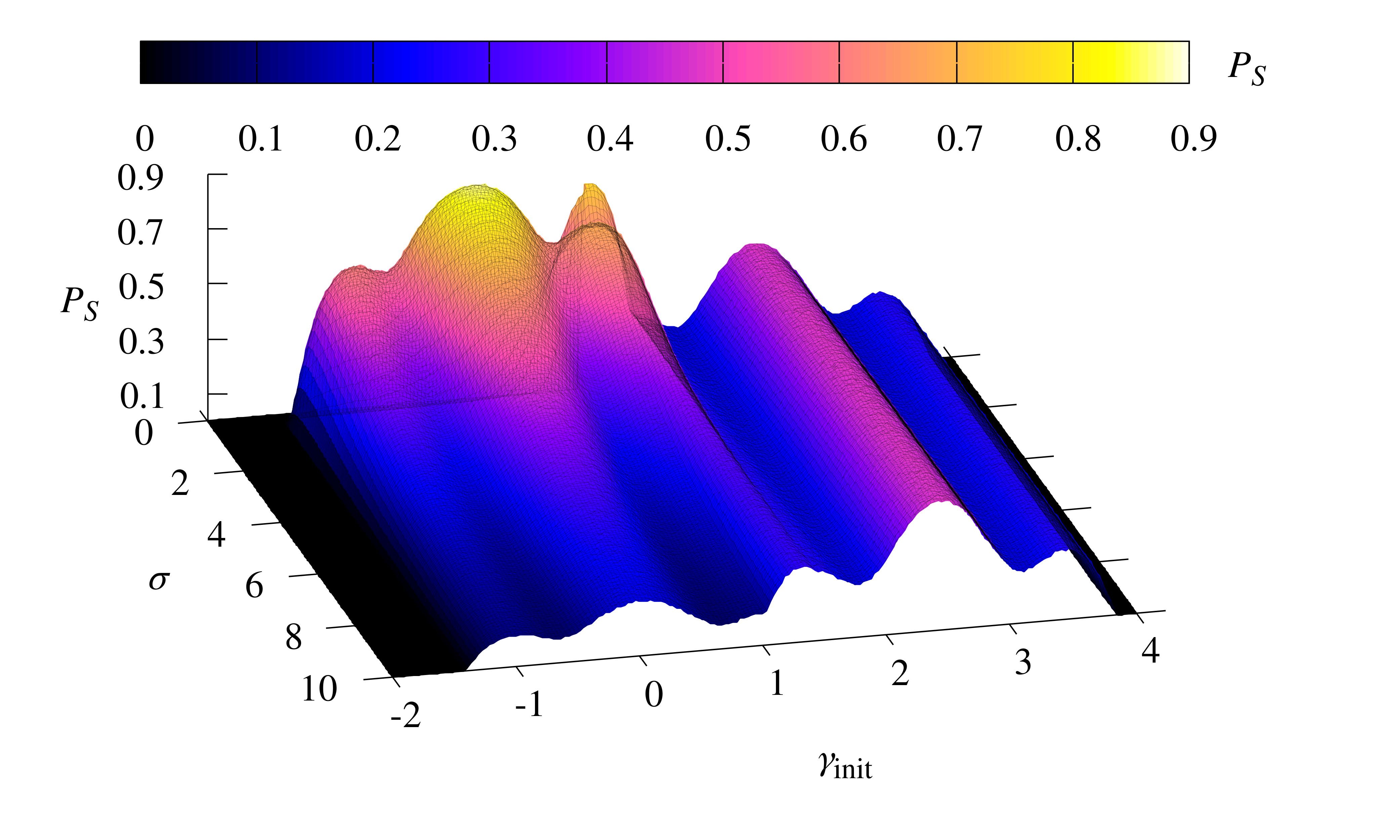}
    \caption{Fraction $P_S$ of outbreaks  where at least one human contracted the disease for different combinations of 
    the initial value of $\gamma_{\text{init}}$ and the mutation rate $\sigma$. Each data point was averaged over 20000 outbreak runs.}
    \label{fig:grid_p}
\end{figure}

First, we wanted to measure how probable it is that at least one human gets infected during an arbitrary outbreak, i.e.,
how probable it is that the disease switches from the animals to the humans.
We therefore simulated 20000 outbreaks for each parameter combination and show the results in \autoref{fig:grid_p}.

Clearly, for a mutation rate with a value of $\sigma=0$, no humans can contract the disease unless the initial value $\gamma_{\text{init}}$
exceeds 1, since the transmission probability to humans is $\lambda^h=0$ below that. 
For slightly larger values of $\sigma$ we observe 6 peaks of $P_S$, which mostly correspond to the peaks of $\lambda^h$ and $\lambda^a$ from 
\autoref{fig:func}. 

Interestingly, the largest peak is at $\gamma_{\text{init}}=0$, where the disease is the most infectious to the animals.
This means, to achieve the highest likelihood of infecting humans, it is more important to first spread well throughout the animal population 
and maximize the number of contacts to the human population, than it is to start with a gene value $\gamma$ that is already able to infect humans.
Note that zoonotic diseases are also often associated with high contact rates of host animals and humans \cite{Slingenbergh2004}

The peak at about $\gamma_{\text{init}}=-1$ is lower than the peak at about $\gamma_{\text{init}}=1$,
even though the corresponding values of $\lambda^a$ are the same.
This makes sense, since for the first case the disease has to mutate more to be able to infect humans.

At $\gamma_{\text{init}}\approx 1.21$ we observe another peak, although at slightly lower mutation rates.
This corresponds to the point where the disease has an equal likelihood of being transmitted to animals and humans.
This peak and all those peaks previously discussed decrease in size for very large mutation rates,
because the subsequent infections will be increasingly dissimilar from 
their parents
and the offspring of a very infectious disease strain are unable to maintain 
this infectiousness.

For values beyond $\gamma_{\text{init}} = \sqrt{2}$ the transmission probability to animals is 0. 
Since the initial patient zero is an animal, this means that the disease can only infect the humans if the initial 
animal infects its owner. Thus the probability $P_S$ becomes independent of $\sigma$, which is visible in the figure.

\begin{figure}[htb]
    \centering
    \includegraphics[width=\linewidth]{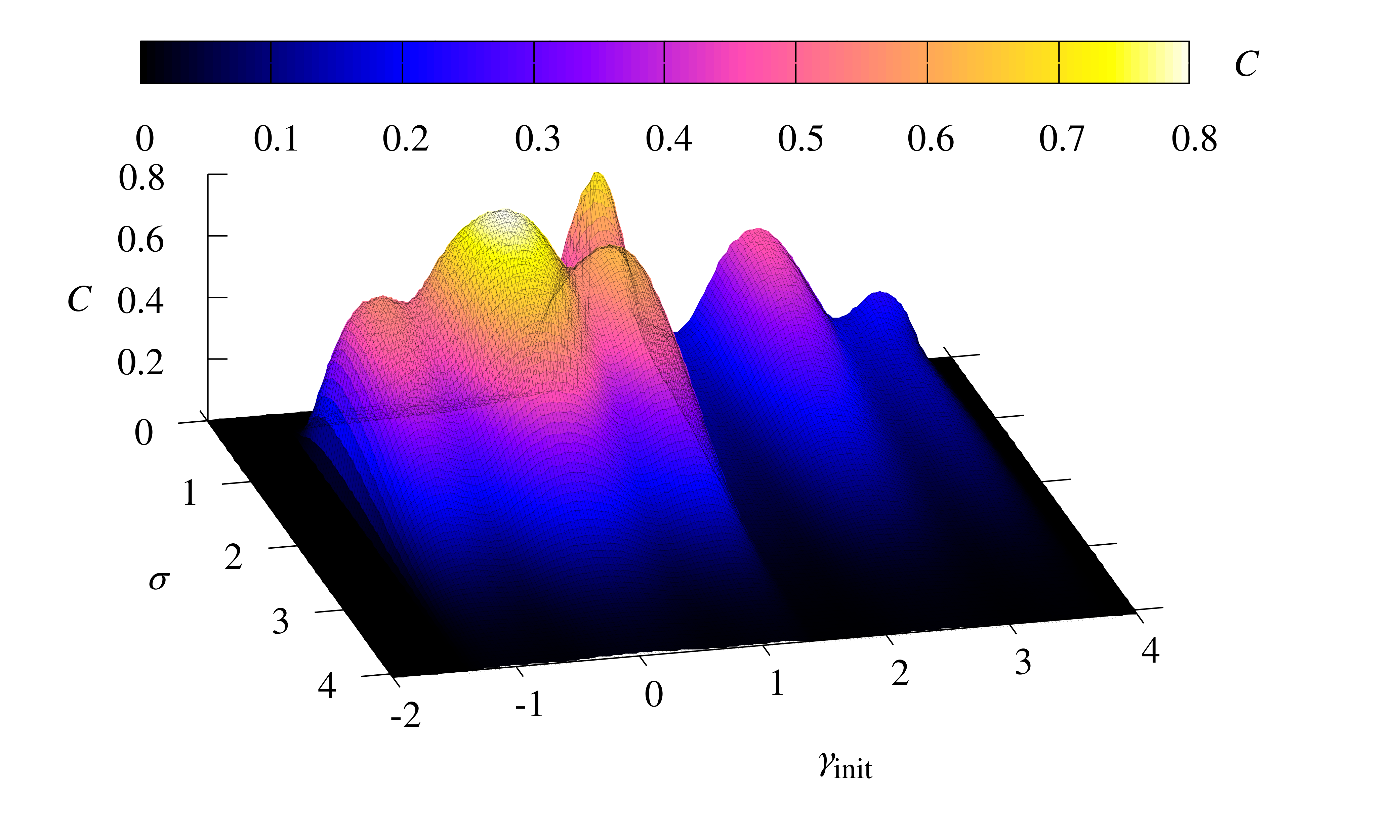}
    \caption{Relative outbreak size in the human population $C$ for different combinations of 
    the initial value of $\gamma_{\text{init}}$ and the mutation rate $\sigma$. 
    Each data point was averaged over 20000 outbreak runs.}
    \label{fig:grid_c}
\end{figure}

Next we looked at the relative outbreak size $C$ in the human population. For this we used the same parameter
as before. In fact, we measured it in the same simulation. The results are displayed in \autoref{fig:grid_c}.
Note that we only display the results up to $\sigma=4$, since $C$ is not distinguishable from 0 beyond that.

Unsurprisingly this plot looks quite similar to the one shown before. This time, however, the 
peak corresponding to the initial value where $\lambda^a=\lambda^h$ is the largest, even though this was not the case for $P_S$,
which shows that, if the outbreaks happen here, they are likely to be more extreme than for the ones
for $\gamma_{\text{init}}=0$.
This is due to the larger initial transmission probability for $\lambda^h$.

If we look at large mutation rates we can see that $C$ decreases monotonically, even for $\gamma_{\text{init}} > \sqrt{2}$.
This was expected, since, even though the mutation rate does not affect the switch probability $P_S$ in this case,
it will affect the outbreak that follows.

\section{Large-deviation simulation}

We next consider the task of precisely measuring the switch probability $P_S$, 
in particular in the case where it is very small.
In this case the typical-event sampling approach becomes unfeasible due 
to the astronomical  amount of samples that this endeavor would require. 
Therefore, we have to turn to the 
large-deviation approach explained in \autoref{sec_ld}.

Note that we always used exactly the same network, i.e., the one we 
already used in the previous section, as discussed in the beginning of the
paper.
To measure one value of $P_S$ for a set of parameters we always performed an entire Wang-Landau simulation, where we used $\ln(f)=10^{-6}$ as termination 
criterion.

\begin{figure}[htb]
    \centering
    \includegraphics[width=\linewidth]{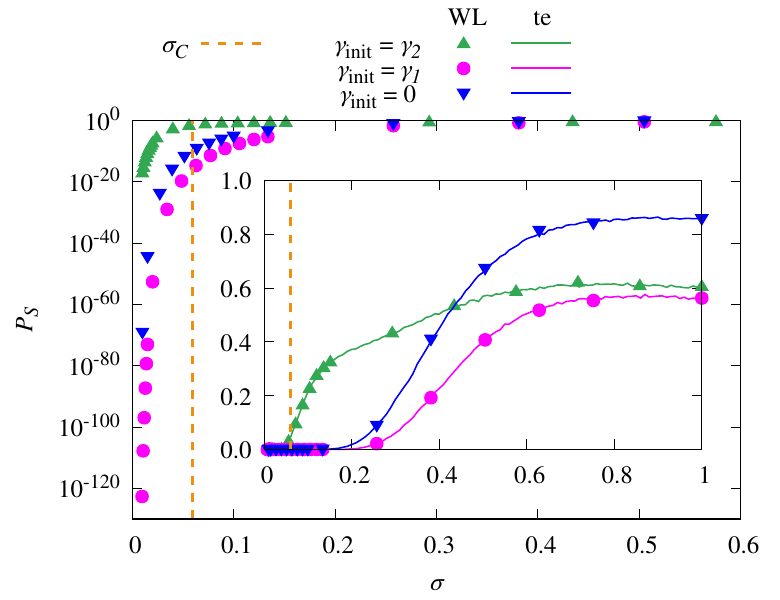}
    \caption{Switch probability $P_S$ measured via Wang-Landau (WL),
shown by symbols,  and typical-event 
    sampling (te), shown by lines,
 for different mutation rates $\sigma$ and initial values $\gamma_{\text{init}}$.
    Note that $\gamma_2=-\gamma_1\approx 0.663$, which is also displayed in \autoref{fig:func}. 
    The dashed line indicates the mutation rate $\sigma_C$ that we use for subsequent simulations. 
    The inset displays the same data until $\sigma=1$ in linear scale}
    \label{SwitchFig1}
\end{figure}

We started our simulations with several distinct initial values of 
$\gamma_{\text{init}} \in \{\gamma_1, 0, \gamma_2\}$ (see \autoref{fig:func}),
where $\gamma_1$ and $\gamma_2$ are the locations where $\lambda^a$ exhibits 
the local minima and $\gamma=0$ corresponds to the global maximum.
For the mutation rates we used various $\sigma$ distributed  in $[0,1]$. 
The results are displayed in \autoref{SwitchFig1}.

Note that we also measured $P_S$ via typical-event sampling, averaging over 30000 samples each,
 in the range where it was easily obtainable .

Looking at the linear range we see the same pattern we have already seen in \autoref{fig:grid_p}, i.e., 
for larger mutation rates the initial ability to spread
within the animal population is more important than starting with a value of $\gamma_{\text{init}}$ that is closer
to being able to infect humans.
For smaller mutation rates, however, this changes and now we observe the largest switch probability $P_S$ for  $\gamma_{\text{init}}=\gamma_2$.
It is also visible here (see also \autoref{fig:grid_p}) that the switch probability starts to decrease again beyond a certain mutation rate, i.e., 
there is an optimal
mutation rate making the switch most likely.
Furthermore, we can observe a very steep decline of the switch probability for small mutation rates and, for $\gamma_{\text{init}}=\gamma_1$
the switch probability becomes smaller than $10^{-120}$.

Overall we can clearly see that the large-deviation approach works very well and enables the calculation of very tiny switch probabilities
with relative ease. 

\begin{figure}[htb]
    \centering
    \includegraphics[width=\linewidth]{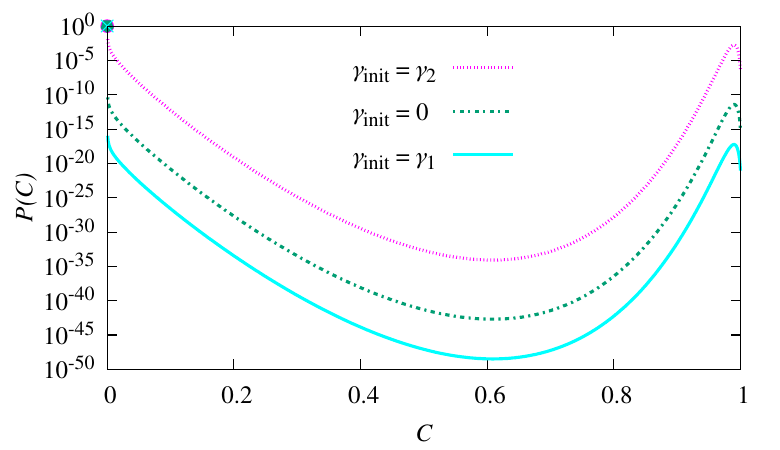}
    \caption{Probability functions $P(C)$ measured with for different initial values $\gamma_{\text{init}}$ 
    with a mutation rate of $\sigma=\sigma_C$. The probability functions are normalized such that the sum over all 1321 bins equals 1.
    Note that the first bin is represented by symbols to highlight the discontinuity.
    }
    \label{figLd1}
\end{figure}

Next we want to investigate the actual size of the outbreak in the human population. 
For this we chose a mutation rate with a value of $\sigma=\sigma_C=0.05939$.
This is in the range where the switch is rather unlikely, but not extremely
unlikely, which, as discussed before, we consider to be realistic. 
We performed  three Wang-Landau simulations, one for each value
$\gamma_{\text{init}} \in \{\gamma_1, 0, \gamma_2\}$. This time we additionally performed entropic sampling afterwards, 
which allowed us to slightly refine the results, although this effect was barely visible at all.
However, since the entropic sampling started with a very good estimate for the probability, i.e., the one obtained with Wang-Landau,
this allows for a rather uniform sampling in the space of different $C$ values, which allow us to simultaneously measure 
other quantities such that we can investigate correlations.
To do this we regularly stored the outbreak trees with additional timing information, which can later be used for the 
analysis.
The resulting pdfs are displayed in \autoref{figLd1}.

Clearly the most probable outcome is $C=0$, i.e., no human infections at all,
and there is a discontinuous drop of probability to the next bin with $C=1$.
For increasing values of $C$ the probability $P(C)$ decreases until $C\approx 0.6$.
Afterwards the probability increases again and the probability for just a few infected 
humans is roughly comparable to the probability of almost all humans contracting the disease.

Interestingly, apart from the different switch probabilities $P_S$, the general shape of the probability functions
for the different $\gamma_{\text{init}}$ seems very similar.
We therefore removed the bin corresponding to $C=0$ and renormalized the results by dividing trough  $1-P_S$.
This confirmed that, apart from the switch probability, the probabilities are exactly the same.
We show a figure for this in the appendix.
As a result, only the outbreak in the animal population is affected by the choice of $\gamma_{\text{init}}$
and the results shown in the following are always for $\gamma_{\text{init}}=0$.

\begin{figure*}[thb]
    \centering
    \includegraphics[width=0.49 \linewidth]{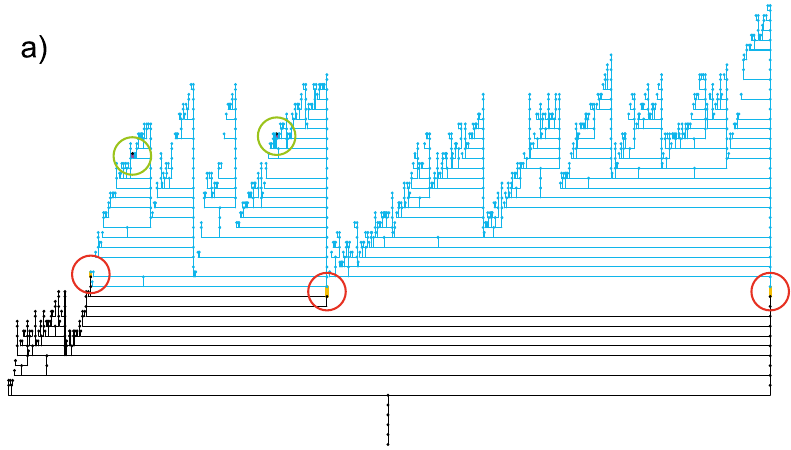} \hfill 
    \includegraphics[width=0.49 \linewidth]{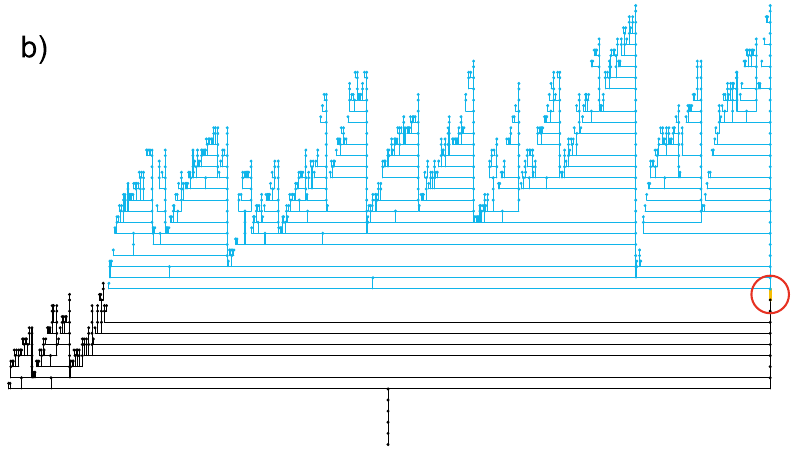} \\
    \includegraphics[width=0.49 \linewidth]{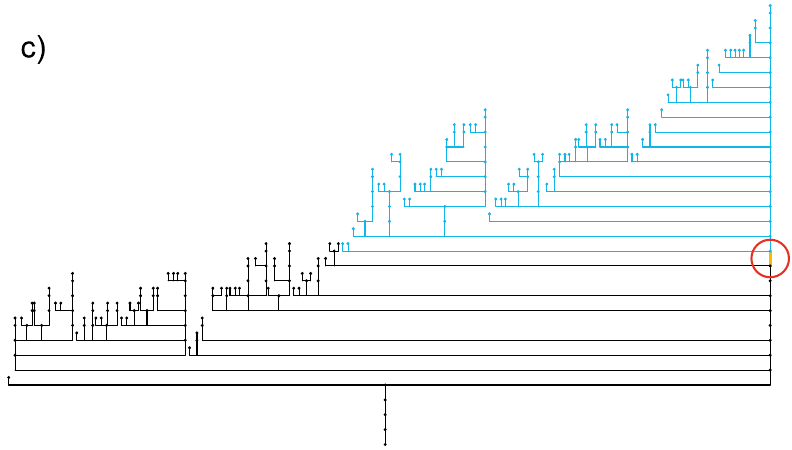}
    \hfill 
    \includegraphics[width=0.49 \linewidth]{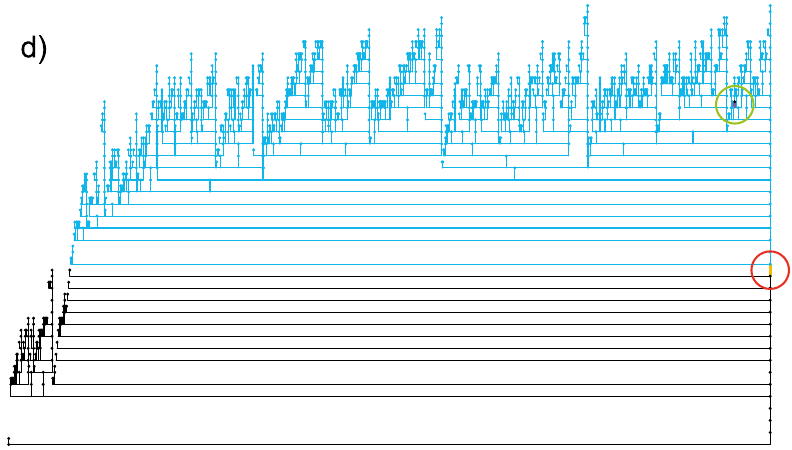}
    \caption{Examples for outbreak trees from the entropic-sampling simulation with $\gamma_{\text{init}}=0$. 
    Here a) and b) are two outbreak trees that correspond to $C\approx 0.6$ (the global minimum in probability),
    c) corresponds to $C=0.1$, while d) corresponds to $C\approx 0.99$ (a local maximum in probability).
    The nodes, represented as dots, display humans (blue) and animals (black).
    Lines represent infection events.
    If a node infected exactly one other node, this other node is plotted 
directly above and connected by a line.
    If it infected several nodes, this is indicated by a horizontal line that 
intersects the node. The children of the node are located above
and connected by vertical lines to the horizontal line.
    The children are sorted from left to right according to the height of the 
sub trees originating from each child.
    Black lines indicate animal-animal transmission, 
blue lines indicate human-human transmission, orange lines (highlighted by red circles) indicate 
animal-human transmission  and lastly purple lines (highlighted by green circles) indicate human-animal 
transmissions.
}
    \label{figTreeExamples}
\end{figure*}

Next we investigate the shape of outbreak trees. 
Some examples can be found in \autoref{figTreeExamples}.
Most trees display one single switch of the pathogen from the animal to the human population. Still, in some cases 
multiple switches occur and if those happened roughly at the same time, 
then it is possible for both switches to result in 
human-network outbreaks of comparable size, like shown in a).
This phenomenon was not specific for $C\approx0.6$ but was instead 
observable for all bins.

Next we want to characterize the outbreak trees. 
Looking at \autoref{figTreeExamples} it seems like for low values 
of $C$ the leafs are located at different 
heights of the tree, while for large values of $C$ the leafs tend to be 
concentrated close to the top of the tree.
To quantify this, we measure  the height of the tree and divided it by the average height of the leafs.
We denote this quantity by $g$.
Small values of $g$ correspond to trees where all leaves exhibits about the
same height, while for larger values of $g$ the leaf heights exhibit
a considerable spread.

\begin{figure}[htb]
    \centering
    \includegraphics[width=\linewidth]{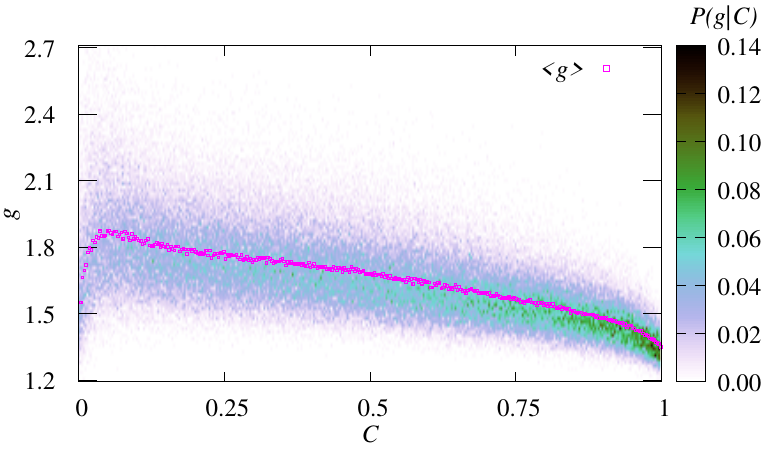}
    \caption{Color coded conditional probability $P(g|C)$. The line
 displays the average $\langle g \rangle$ as a function of $C$.}
    \label{figG}
\end{figure}

The results for the conditional probability $P(g|C)$ are shown 
in \autoref{figG}.
We see that the average $\langle g \rangle=\int dg\, g P(g|C)$ 
peaks around $C\approx0.05$ with a value of about 1.85.
For $C<0.05$ the value of $g$ tends to be lower, which makes sense. 
The disease has to mutate to be able to infect the humans and as visible in \autoref{figTreeExamples}
the switch of the pathogen tends to happen at or at least close to the 
top of the outbreak tree, i.e., most levels of the tree describe
the animal dynamics.
For $C>0.05$, however, the average $\langle g \rangle$ 
decreases monotonously with $C$, i.e., the bulk of the 
leafs is located closer and closer to the top of the tree. This 
confirms the impression already obtained when looking at the sample trees,
that large outbreaks are characterized by broad infections
fronts, i.e., many independent infection events take place at the same time.

Although our model is quite limited by encoding the entire gene of the 
disease by just one value, $\gamma$, one can, in a restricted way,
identify ``variants'' of the disease in the following way.
We start at any human $i$ and just treat him as the origin of a new 
variant. Then we consider the subtree consisting only of node $i$ and 
its descendants and iteratively follow every path to the leafs. 
For each of these paths $p$ we keep track of the minimum $\gamma_{\min}^p$ 
and maximum $\gamma_{\max}^p$ 
of  the values of  $\gamma$ that is encountered. 
If the encountered fluctuations along a considered path 
are larger than some pre-chosen threshold $\Delta \gamma$, i.e., if 
 $\gamma_{\max}^p - \gamma_{\min}^p > \Delta \gamma$, then we conclude 
that the ``existence'' of the variant has ended and stop the path
at the corresponding node.
 We do this for all possible infection paths, which is simply
achieved
by a recursive function without the need to enumerate
all paths, and count the number $R_i$ of 
nodes that are part of this restricted tree starting at human $i$.
The \emph{reach} $R_i$ is a kind of topological measure of the impact
of the variant starting at node $i$.

\begin{figure}[htb]
    \centering
    \includegraphics[width=\linewidth]{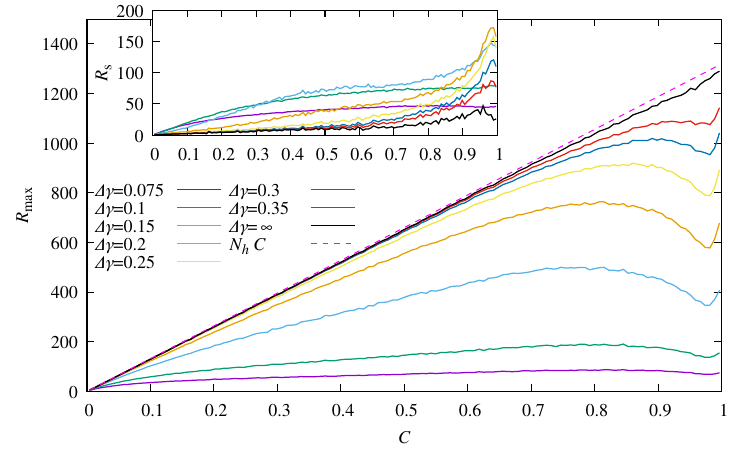}
    \caption{Averaged largest reach $R_{\max}$ for different values of $\Delta \gamma$ as a function of C.
    The dashed line indicates $N_h C$, i.e., the total number of humans that were infected.
    The inset shows the averaged second largest reach $R_s$. }
    \label{figMaxReach}
\end{figure}

 Thus, we can now define the \emph{maximum reach} as $R_{\max} = \max_i R_i$.
 In a similar fashion we calculate the second-largest reach $R_s$. We do the same calculation as before, but first we remove 
 all nodes that contribute to $R_{\max}$ from the outbreak tree.
 We display the results of the averaged quantities for different values of 
$\Delta \gamma$ in \autoref{figMaxReach}. 

For $\Delta \gamma = \infty$ the largest reach results from a
 human that was infected by an animal,
since all of their descendants will be counted for $R_{\max}$.
If all outbreak trees exhibited exactly one switch to the human population, 
this would always equal the 
total number of human infections, i.e., it would be equal to $N_h C$.
Since we sometimes observe more than one switch to the human population, 
this quantity is slightly below this bound.

This effect is even more visible in the second-largest reach $R_s$, since 
it would be zero if all humans contributed 
to the largest reach. It is, however, actually always larger than zero, 
although for $R_s<0.1$ this is hard to see in the plot.

By decreasing $\Delta \gamma$ we see a local minimum appearing for $R_{\max}$ near $C=0.99$, 
which corresponds to a maximum of $R_s$.
Interestingly this also corresponds to a local maximum of $P(C)$,
see \autoref{figLd1}. From investigating sample outbreak trees
it appears that here  separate branches occur which exhibit 
high infection probability $\lambda^h$ independently of one another.
On the other hand for $C=1$ one again observes  relatively higher values of 
$R_{\max}$   and lower values of $R_s$, which indicates that here 
one single rather infectious strain was able to evolve.

For each outbreak tree we can compute the maximum value of $\lambda^h$ that was achieved. Looking at the average $\left<\lambda^h_{\max}\right>$ 
 as a function of $C$ (plot in the appendix) we observe that around $C\approx 0.95$
 the disease manages to overcome the barrier that results from the 
local minimum of $\lambda^h$ of 0.0396 which is
visible in \autoref{fig:func}.

If we decrease $\Delta \gamma$ even further, then the curves for $R_{\max}$ become quite flat and are of similar magnitude 
as $R_s$. This is similar to the case of studying the largest and second largest
component of random-graph percolation \cite{hartmann2005, Newman2010} 
and might indicate that $R_{\max}$ switched from an
extensive to an intensive quantity. But one would have to measure this with 
multiple system sizes and perform a finite-size analysis until
one can conclude that a percolation-like phenomenon is present.

\begin{figure}[htb]
    \centering
    \includegraphics[width=\linewidth]{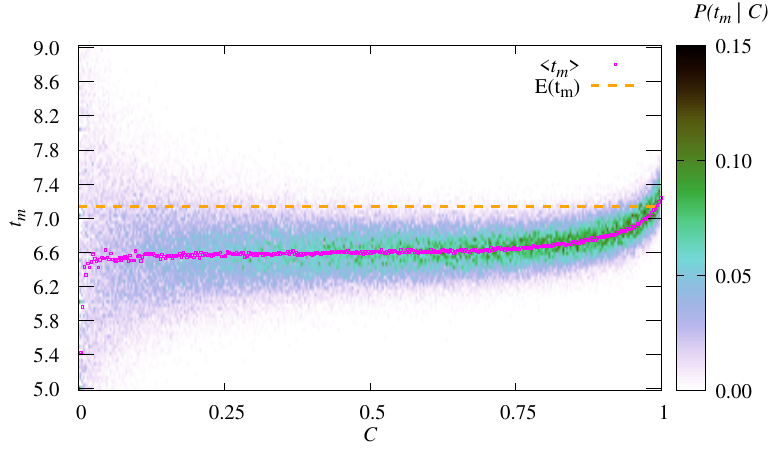}
    \caption{Color coded conditional probabilities $P(t_m|C)$. We additionally display the average of $\left<t_m\right>$ as a function of $C$.}
    \label{heatmapRec}
\end{figure}

Next we take a look at how long the humans take to recover
and the relation to the ourbreak dynamics. 
For each outbreak we calculated the mean recovery time $t_m$ 
that the humans took to recover, where the mean is taken over all
infected humans. This will change from outbreak to outbreak and 
results in the conditional probability $P(t_m|C)$, see
 \autoref{heatmapRec}.
We also included the averaged mean recovery time 
$\left<t_m\right> = \int dt_m\,t_m P(t_m|C)$ 
conditioned to $C$. Furthermore, we show 
the total expected recovery time 
\begin{equation}
    E(t_m)=\sum_{n=1}^\infty n(1-\mu)^{n-1} \mu  = \frac{1}{\mu}\approx 7.14~,
\label{eq:E:tm}
\end{equation}
as a dashed horizontal line in the plot.
At first glance it might seem strange that the average recovery time lies mostly beneath this  expected value. But one has to take into account that
an outbreak with a fraction $C$ of humans contributes proportionally to $C$
and to $P(C)$ to the statistics, thus 
\begin{equation}
    E(t_m)=\frac{\sum_C P(C) C E(t_m|C)}{\sum_C P(C) C}~
    \label{eqChecking}
\end{equation}
should hold. This average is dominated from regions where $CP(C)$ is large,
which here is for $C$ near 0.95. 
And indeed, plugging in our numerical results we obtain
 approximately 7.14, which fits \autoref{eq:E:tm}.

For small values of $C$ the mean recovery time scatters quite a lot, which makes sense given that 
this quantity is obtained from averaging the recovery time of all infected humans and for small values of $C$ only very few humans 
become infected.
Still, on average the values close to $C= 0$ have noticeably shorter recovery times and there is a 
steep increase in the first few datapoints of 
$\left<t_m\right>$. This makes sense, as a very fast recovery of the 
first infected humans makes it less likely that the disease is 
transmitted further.
Thus, unusual small outbreaks arise due to unusual quick recoveries.

For larger values of $C$ the mean recovery time becomes more and more concentrated around the average $\left<t_m\right>$
which is close to a value of $6.6$ with a slight incline that is barely noticeable.
At about $C=0.8$ the slope becomes steeper and $\left<t_m\right>$ increases noticeably
and peaks at about $\left<t_m\right>=7.3$ for $C=1$.
So outbreaks that reach every single human are characterized by having a larger recovery time on average, which is also reasonable.
It is worth mentioning that the mean recovery of the animal population does 
not seem to correlate with $C$ at all (not shown).

\begin{figure}[htb]
    \centering
    \includegraphics[width=\linewidth]{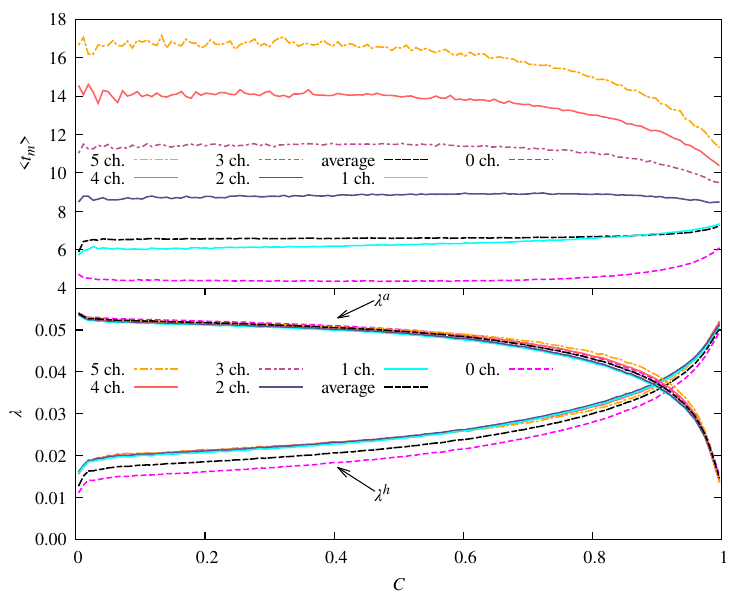}
    \caption{(top) 
Mean recovery time $\left<t_m\right>$ averaged only over nodes 
with a given number of children (ch.) and conditioned to a value of $C$,
    as a function of $C$. 
We also included the average independent of the number of children, i.e., 
the one that was already displayed in \autoref{heatmapRec}.
    (bottom) Averaged transmission rates $\lambda^h$ and $\lambda^a$, again 
averaged over nodes with a specific number of children,
    as a function of $C$.
    Again, we include the averages calculated independent of the 
number of children for reference.
    }
    \label{ChidrenRecLamFig}
\end{figure}

Next we explore the influence of the recovery even a bit further. 
For each outbreak tree we sort the human nodes of the outbreak tree by 
the number of children.
Then we were able to calculate the mean recovery time for a given 
number of children. In general it can be expected that
longer recovery times lead to a higher number of children.
This can be indeed observed in the top of \autoref{ChidrenRecLamFig}
where the average recovery time conditioned to $C$ and to the
number of children is shown as function of $C$. The corresponding
plot for the observed transmission rates is shown in the bottom plot.

One can clearly see that the number of children is correlated with the 
recovery time. 
Nodes with extremely short recovery times tend to have no children, while 
nodes with more children tend to have longer recovery times.

If we look at increasing values of $C$ we can see that the recovery time of nodes with 3 or more children tends
to become shorter, the recovery time of nodes with 1 or fewer children tend to become longer,
while the recovery time of nodes with 2 children are, compared to the others, mostly unaffected.
This is a result of two different mechanisms that are at play here. 
At the later stages of the disease, which are more relevant for
larger values of $C$, less and less susceptible nodes remain and thus 
a longer recovery time is required to directly infect the same number of nodes.
On the contrary, diseases that have a higher transmission 
probability $\lambda^h$ tend to have 
more offspring and thus the disease becomes more infectious. 
This effect can be seen at the bottom of 
\autoref{ChidrenRecLamFig} where higher values of $C$ are correlated with larger values of $\lambda^h$.
A more infectious disease needs less time to infect the same number of 
neighbors.
For the nodes with 2 children, these effects seem to roughly balance out. 
Meanwhile, for nodes with several children the effect of increased transmission probability dominates, 
whereas for nodes with few children, the effect of the decreasing 
number susceptible nodes prevails.

We also observe that nodes without children, i.e., leafs, tend to have 
a smaller value of $\lambda^h$ than their counterparts, while 
nodes with any other number of children display 
very similar transmission probabilities.

Also, on average, the disease is more infectious to animals than to humans,
reminding us of its animal origin. This  only reverses for large human pandemics
with $C>0.909$.
Clearly, this is a result of the chosen functions $\lambda(\gamma)$
and might result in more noticeable effects if we had used a 
larger animal population.
Here, however, it is quite likely that most animals cannot contract 
the disease any longer at the time, when the first human gets infected.

Next we investigate a related property. We are interested to see how the average transmission probability $\left<\lambda\right>$ measured at 
the human node which  resulted
in most offspring behaves as a function of $C$. This human was necessarily 
infected by an animal,
because otherwise its parent would have been responsible for
more offspring than itself.
Note that this human does not have to be the first human that got infected, 
since multiple switches to the 
human population are possible and their infection trees are separate.

\begin{figure}[htb]
    \centering
    \includegraphics[width=\linewidth]{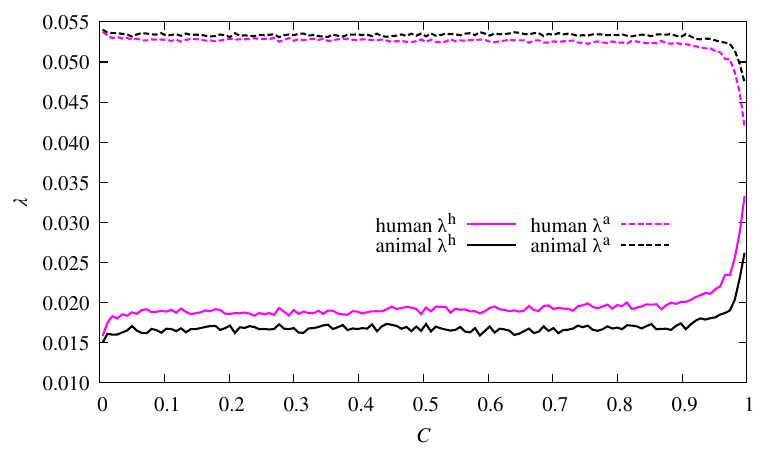}
    \caption{Transmission probability $\lambda^h$ to humans  
and transmission probability $\lambda^a$ to animals  at  the human 
    node $i$ with the most pathogen offspring and the animal that 
infected node $i$, both as a function of $C$.
    Note that we display $\lambda^a$ only for the sake of completeness.}
    \label{LambdaHumanMostOffspring}
\end{figure}

We display the result in \autoref{LambdaHumanMostOffspring}.
Clearly the disease tends to exhibit considerable mutations for 
transmission from animal to human, which is visible by the small but clear
separation between the two $\lambda^h(C)$ curves.
In our statistics the cases will dominate where this human is more 
infectious to other humans than the corresponding animal, because
we almost exclusively (except for $C N^h=1$ or other very low value of $C N^h$ with multiple 
host switching events)
measure outbreak trees where this human transmits the disease further. 
This corresponds well to the fact that $\lambda^h(C)$ for the human is
above the curve for animals. 

Looking at the dependence as function of $C$,
up to $C=0.9$ the transmission probability of human and animal, respectively,
 do not  change much and thus 
are not a good indicator for estimating the size of the outbreak. 
However, for $C>0.9$ we see a clear rise of $\lambda^h$ in the human and 
even in its animal,
which means that the very large outbreaks where almost every human 
gets infected are, on average,
originating from a disease that was able to already obtain a higher 
infectiousness to humans in the animal population. This  shows
that controlling zoonoses within animal populations has a benefit
for the human population as well.

Finally, we want to take a look at the mutation events that occur within the human population.
As explained in \autoref{sir_model_sec}, the disease mutates, more or less,
 each time it is transmitted. 
Since this mutation is drawn from a Gaussian with mean 0, on average $50\%$ of the mutations 
should lead to a reduction of the $\gamma$ value and vise versa, although this may not hold true if we constrain the system 
to specific values of $C$. 

Anyhow, we are more interested in the transmission probability $\lambda^h$ 
to humans, since this is what ultimately affects the 
spread of disease within the human population.
For this purpose, we now only consider transmissions that occurred 
between humans. Based on the outbreak trees, we computed the fraction 
of transmissions that lead to a 
reduction of an arbitrary quantity $q$, which we will call 
negative mutation fraction of $q$ and denote it by 
$f_{-}^q$.

\begin{figure}[htb]
    \centering
    \includegraphics[width=\linewidth]{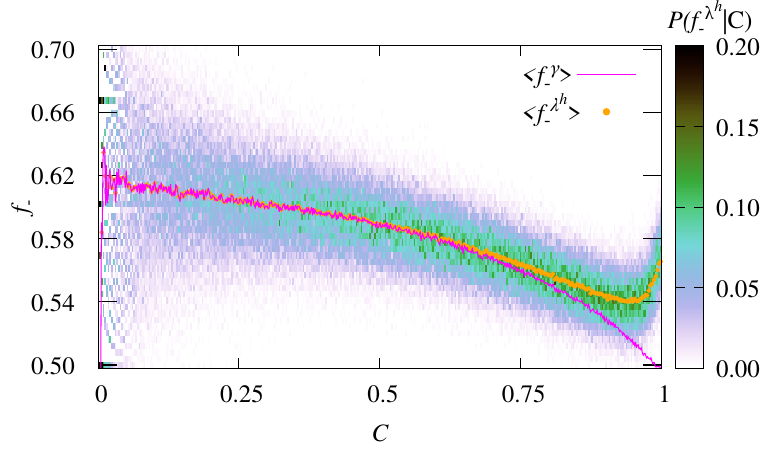}
    \caption{Color coded conditional probability $P(f_{-}^{\lambda^{h}}|C)$.
    We also display the average $\left<f_{-}^{\lambda^{h}}\right>$ and the 
average $\left<f_{-}^{\gamma}\right>$  as a function of $C$.}
    \label{neg_lambda_mut}
\end{figure}

Of special interest is the negative mutation fraction $\lambda^h$, 
since these are the mutations that led to a reduction of 
the transmission probability. We display the results for this quantity 
in \autoref{neg_lambda_mut}.
Additionally, we show the average $\left<f_{-}^{\gamma}\right>$.

Apart from a few outliers for very small values of $C$, 
which are not displayed, the negative mutation fraction of the transmission probability $\lambda^h$
 always exhibits values larger than 0.5. 
At first glance this may seem surprising.
However, we only expect an overall average fraction of $0.5$ for the fraction of negative changes for $\gamma$, 
not for the fraction of negative  changes
for $\lambda^h$,
since the function connecting the two is not monotonous.
And if we take the probability into account and do a calculation analogously to \autoref{eqChecking} we end up with 
an expected value that can be calculated via
\begin{equation}
    E(f_{-}^{\gamma})=\frac{\sum_C P(C) C E(f_{-}^{\gamma}|C)}{\sum_C P(C) C}~.
    \label{eqChecking2}
\end{equation}
If we plug in our numerical results we get $E(f_{-}^{\gamma})=0.5$, exactly as expected.
Note that this calculation ignores that we have at most $C (N^h - 1)$ mutations from human to human transmissions, 
because the first human gets infected through an animal, but this
effect is negligible. Also, as we have seen, there can be more than 
one human that gets infected from an animal in an outbreak, but this fraction was always very small 
in our results, so \autoref{eqChecking2} is a good approximation.

Coming back to \autoref{neg_lambda_mut} we can see that $\left<f_{-}^{\lambda^{h}}\right>$ and $\left<f_{-}^{\gamma}\right>$
are almost identical for $C<0.7$. 
At around $C\approx 0.7$ they start to diverge and for $C\approx 0.95$
the negative mutation fraction of $\lambda^h$ even starts to increase, even though the fraction 
for $\gamma$ continues to decrease.

There are two reasons for that. Firstly, if $\lambda^h$ reaches the first local maximum (see \autoref{fig:func})
then any mutation will decrease the transmission probability, which is why the quantities $\left<f_{-}^{\lambda^{h}}\right>$ and $\left<f_{-}^{\gamma}\right>$
 start diverge at around $C \approx 0.7$. 
Secondly, 
as we mentioned earlier at around $C\approx 0.95$ we start to see values of $\lambda^h$ that exceed the first local maximum
(see appendix). Clearly, this requires to go through the local minimum, 
i.e., many mutations that decrease the transmission probability must
be present in the outbreak tree,
which explains the results visible in \autoref{neg_lambda_mut}.

\section{Summary and outlook}

With the presented model we study the spread of a disease in 
a combined animal-human network for a pathogen which is characterized by a,
yet simple, fitness landscape. As explained, only diseases
are still evolutionary relevant, where the animal-human host switch exhibits
a very small transmission probability per animal-human contact.

While previous studies mostly analyzed the danger of host switching events by applying a 
branching-process that starts with the first infected humans,
we were able to model the entire process, starting from a disease which is not able to infect humans.
This disease changes through mutations and results in a disease that 
is able to cause an epidemic outbreak in the human population.

Using large deviation techniques we were able to numerically cope with the
very small  probabilities of 
a host switching event occurring for a given disease.
Note that our approach could be used for any other functional relation between the value of $\gamma$ and the 
transmission probabilities. In fact, this approach can also be extended to less trivial functions.
In particular one could consider multi-dimensional fitness landscapes, 
e.g.,  introduce additional gene variables $\alpha, \beta, …$ and let the 
transmission probability be a function $\lambda=\lambda(\alpha, \beta, \gamma, …)$.
Ideally one might be even be able to infer an approximation from the 
genome of actual real-world diseases.

Furthermore we were able to calculate the complete 
probability density function $P(C)$ of the cumulative fraction of infected humans 
that characterizes the outbreak.

We are able to analyze the entire outbreak trees that capture the outbreak dynamics. It is worth mentioning that, while the large-deviation simulation is certainly 
computationally expensive, the successive analysis of the stored 
outbreak trees is quite cheap and therefore fast.
Note that storing the trees also allows for the analysis of other quantities 
that one does have in mind when performing the large-deviation
simulations.

By measuring the correlations with other quantities we were able to see that 
outbreaks that only affect a fraction of the human population are characterized by 
faster recoveries as compared to outbreaks that reach the entire population.

Also, even if the host switching event itself is quite improbable, once the disease manages to mutate such that 
one host switching event occurs, it is quite probable that further events occur.
Given that the typical size of populations in the real world is much larger and, in contrast to the applied SIR model,
might allow reinfections, especially given that the disease mutates, 
diseases that have shown host switching events are of special concern.

On the other hand, at least in this simplified model, a host switching event is the result of the disease gradually becoming 
more infectious to humans and not characterized by a huge single mutation.

Overall we have shown how large-deviation methods can be applied as an important tool for 
understanding host switching events and further studies using the same methods are likely to 
be very useful for understanding and therefore an aid in preventing host switching 
events in specific pathogens. Many different research directions,
for various fitness landscapes, network types, or more complex
disease propagation models, can be considered in this way.

\begin{acknowledgments}

We thank Yvonne Feld for helping with figure 6.

Yannick Feld has been financially supported by the German Academic Scholarship Foundation (Studien\-stiftung des Deutschen Volkes).

The simulations were performed at the HPC Cluster CARL, located at the University of Oldenburg (Germany) and funded by the DFG through its Major Research Instrumentation Program (INST 184/157-1 FUGG) and the Ministry of Science and Culture (MWK)
of the Lower Saxony State.

This work also used the Scientific Compute Cluster at GWDG, the joint data center of Max Planck Society for the Advancement of Science (MPG) and University of Göttingen.

The source code for this study was written in Rust and can be found on Github at \url{https://github.com/Pardoxa/sir_animal}.

\end{acknowledgments}

\appendix
\section{Shifted pdfs}

\begin{figure}[htb]
    \centering
    \includegraphics[width=\linewidth]{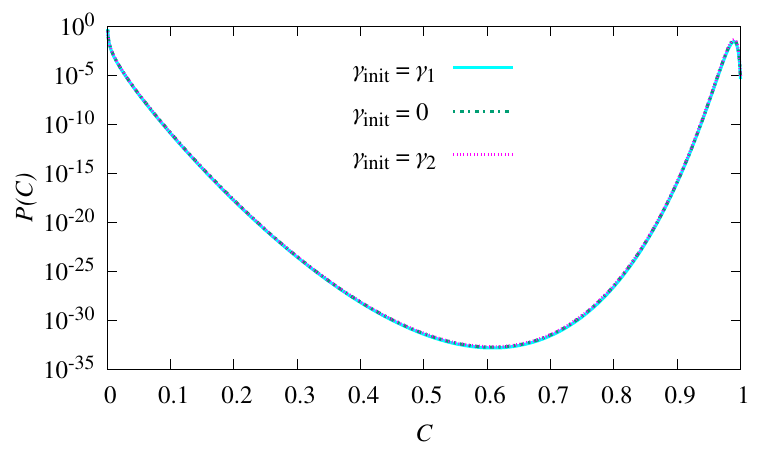}
    \caption{Probability functions $P(C)$ measured with for different initial values $\gamma_{\text{init}}$ 
    with a mutation rate of $\sigma=\sigma_C$. The bin corresponding to $C=0$ was removed 
    and the remaining bins were renomalized such that their sum is 1.}
    \label{figLdShifted}
\end{figure}

In \autoref{figLdShifted} we show the shifted probability density functions mentioned in the main text.
The different probability density functions match exactly.

\section{Maximal value of $\lambda^h$}

\begin{figure}[htb]
    \centering
    \includegraphics[width=\linewidth]{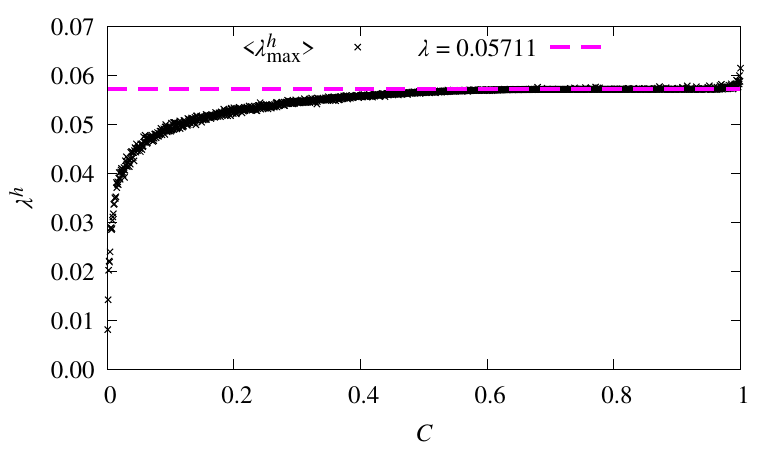}
    \caption{Average $\left<\lambda^h_{\max}\right>$ of the maximal value of $\lambda^h$ encountered in the outbreak trees as a function of $C$.
    The dashed line indicates the value of $\lambda$ that corresponds to the first local maximum of the function $\lambda(\gamma)$.
    }
    \label{lambdaMax}
\end{figure}

In \autoref{lambdaMax} we show the maximal value of $\lambda^h$ that was reached during the outbreak simulations as a function of $C$.
Clearly, the infection probability of most outbreaks is limited by the first local maximum, while the 
local minimum next to it can be seen as some sort of barrier. 
However, some outbreaks are able to pass this barrier and achieve very large values of $C$. 

\bibliography{bib.bib}
\end{document}